\documentclass[referee,sn-basic]{sn-jnl}

\usepackage{graphicx}%
\usepackage{multirow}%
\usepackage{amsmath,amssymb,amsfonts}%
\usepackage{amsthm}%
\usepackage{mathrsfs}%
\usepackage[title]{appendix}%
\usepackage{xcolor}%
\usepackage{textcomp}%
\usepackage{manyfoot}%
\usepackage{booktabs}%
\usepackage{algorithm}%
\usepackage{algorithmicx}%
\usepackage{algpseudocode}%
\usepackage{listings}%
\usepackage{enumerate}%

\newcommand{\wh}{\widehat}
\newcommand{\wt}{\widetilde}
\newcommand{\ve}{\varepsilon}

\newcommand{\bB}{{\mathbf B}}

\newcommand{\bS}{{\mathbf S}}

\newcommand{\bX}{{\mathbf X}}
\newcommand{\bY}{{\mathbf Y}}
\newcommand{\bZ}{{\mathbf Z}}

\newcommand{\bff}{{\mathbf f}}

\newcommand{\bbeta}  {\boldsymbol{\beta}}

\newcommand{\bOmega}{\boldsymbol{\Omega}}
\newcommand{\bomega}{\boldsymbol{\omega}}

\newcommand{\bgamma}{\boldsymbol{\gamma}}

\newcommand{\btheta} {\boldsymbol{\theta}}
\newcommand{\bxi} {\boldsymbol{\xi}}

\newcommand{\bzero}{{\mathbf 0}}

\theoremstyle{thmstyleone}%
\newtheorem{theorem}{Theorem}
\theoremstyle{thmstyletwo}%
\newtheorem{remark}{Remark}%

\theoremstyle{thmstylethree}%
\newtheorem{definition}{Definition}%
\newtheorem{condition}[definition]{Condition}%

\begin{document}

\title[Cross-validation Model Averaging]{Model Averaging by Cross-validation for Partial Linear Functional Additive Models}


\author[1]{\fnm{Shishi} \sur{Liu}}\email{liushishi@hdu.edu.cn}

\author*[2]{\fnm{Jingxiao} \sur{Zhang}}\email{zhjxiaoruc@163.com}


\affil[1]{\orgdiv{School of Economics}, \orgname{Hangzhou Dianzi University}, \orgaddress{\city{Hangzhou}, \postcode{310018}, \country{China}}}

\affil*[2]{\orgdiv{Center for Applied Statistics, School of Statistics}, \orgname{Renmin University of China}, \orgaddress{\city{Beijing}, \postcode{100872}, \country{China}}}


\abstract{In this paper, we propose a model averaging approach for addressing model uncertainty in the context of partial linear functional additive models. These models are designed to describe the relation between a response and mixed-types of predictors by incorporating both the parametric effect of scalar variables and the additive effect of a functional variable. The proposed model averaging scheme assigns weights to candidate models based on the minimization of a multi-fold cross-validation criterion. Furthermore, we establish the asymptotic optimality of the resulting estimator in terms of achieving the lowest possible square prediction error loss under model misspecification. Extensive simulation studies and an application to a near infrared spectra dataset are presented to support and illustrate our method.}




\maketitle

\section{Introduction}
\label{intro}
The increasing prevalence of large datasets with continuous recording in domains such as meteorology, biology, and medical science has highlighted the significance of functional data analysis (FDA) as a powerful tool for modeling variables defined on a continuum, such as time and space. \cite{ramsaysilverman2005} provides a comprehensive introduction to FDA tools for diverse problems, including functional principal component analysis (FPCA), correlation analysis, clustering, classification, and regression. 

Regression models involving functional variables play a major role in the FDA literature. One of the most well studied regression models is the functional linear model, which incorporates the effect of a functional predictor by an integral of the product of the functional variable and its corresponding coefficient function. Extensive researches have been dedicated to functional linear models, as evidenced by the works of \cite{cardotetal1999,cardotetal2003,yaoetal2005,caihall2006,caiyuan2012}. Additionally, researchers have extended the framework of functional linear models to generalized functional linear models to deal with generalized response variables, see \cite{james2002,mullerstadtmuller2005,lietal2010}, among others. All these studies assume specific forms of the regression model, leading to a characterization as functional parametric regression models, as described by \cite{ferratyvieu2006}.

However, the assumption of a linear relationship between the response and the functional predictor may be too restrictive to capture characteristics of real-world data. Consequently, researchers have focused on functional nonparametric regression models, which do not impose structural constraints on the regression function and have the ability to detect nonlinear relationship. Various estimation approaches have been proposed, including kernel-type estimator \citep{ferratyvieu2002,chagnyroche2016}, local linear estimation procedures \citep{baillograne2009,bojetal2010}, and k-nearest neighbour (kNN) estimators \citep{burbaetal2009,karaetal2017}. Functional nonparametric regression models can output less biased estimates than functional linear regression when nonlinear relationships do exist in the data. Nevertheless, it is widely acknowledged that functional nonparametric modeling suffers from the ``curse of dimensionality'' both theoretically and practically \citep{geenen2011}. To address this issue, semi-parametric regression models for functional data have been widely proposed, for example, functional single index models \citep{chenetal2011,jiangwang2011,sangcao2020}, functional additive models \citep{mulleryao2008,fanetal2015}, functional quadratic models \citep{yaomuller2010,sunwang2020}, which aim to preserve the modeling flexibility of functional nonparametric regression while mitigating the problems related to the ``curse of dimensionality''.

All these aforementioned functional regression models mainly focus on the relationship between a response variable and a functional predictor. However, in practical scenarios, data collection often involves not only functional predictors but also scalar predictors that impact the response variable. This type of data, referred to as mixed data or hybrid data in \cite{ramsaysilverman2005} has received considerable attention in the literature. For instance, \cite{aneirosvieu2006} introduced a semi-functional partial linear model that combines nonparametric modeling of the functional variable with linear modeling of the scalar variables. \cite{wangetal2016} proposed functional partial linear single-index models, which treat scalar predictors as linear part while accommodating the functional predictor through a single-index component. Furthermore, \cite{kongetal2016} and \cite{yuetal2016} developed partial functional linear regression and partial functional linear quantile regression models, respectively, offering more flexibility and robustness. These studies demonstrate the importance of the partial linear structure and its widespread application in addressing problems with mixed data. Building on these foundations, \cite{wongetal2019} enhanced the modeling flexibility of the functional predictor by introducing an additive effect based on functional principal component (FPC) scores, leading to a clear improvement in regression fitting and prediction. Motivated by these advancements, our work aims to leverage the popular partial linear structure to effectively handle mixed data and adopt a flexible semi-parametric approach to model the functional predictor in an additive form, leading to our focus on partial linear functional additive models. 

In regression analysis, including partial linear functional additive models, the uncertainty of determining which predictors or functional components to include is a common challenge. All of the estimation approaches proposed in above works address this model uncertainty by employing model selection strategies and make predictions based on a single selected model. However, it has been acknowledged that the uncertainty introduced by the selection process is usually ignored by model selection methods, leading to overconfident estimators that underestimate the associated risk \citep{clydegeorge2004,claeskenshjort2008}. To mitigate this issue, we propose to use a model averaging technique, which combines multiple candidate models with proper weights to produce a more robust result. Model averaging offers an alternative approach to tackling model uncertain and has been extensively studied in the literature, including Bayesian model averaging and frequentist model averaging methods. While Bayesian model averaging \citep{hoetingetal1999} works well when suitable prior distributions for the weights can be specified, the more recently developed frequentist paradigm for model averaging is preferred in cases where prior knowledge is lacking.

A rich body of frequentist model averaging methods are available in the literature. One category of these methods, known as combination for adaption methods, aims to approach the performance of the best single model, see \cite{yang2001}, for example. Another category of combination methods is optimal model averaging methods, which are designed to outperform any single model. Several notable studies include information criterion-based weighting scheme \citep{bucklandetal1997}, Mallows model averaging \citep{hansen2007,wanetal2010}, optimal mean squared error averaging \citep{liangetal2011}, Kullback-Leibler loss-based model averaging \citep{zhangetal2015,fangetal2022,zouetal2022}, jackknife model averaging \citep{hansenracine2012}, and cross-validation model averaging \citep{zhangetal2013,chenghansen2015,gaoetal2016,zhangliu2023}, each tailored to specific non-functional regression models or non-functional data scenarios. 

Recently, there has been growing interest in applying model averaging techniques to functional regression models. \cite{zhangetal2018} developed a model averaging method for function-on-function regression models, where both the response and the predictor variables are of functional type. This approach involved selecting optimal weights by minimizing a multi-fold cross-validation criterion. \cite{zhuetal2018} focused on partial linear functional linear models and introduced an optimal model averaging estimator based on a Mallows-type criterion, derived from an unbiased estimate of mean squared error risk. \cite{zhangzou2020} extended the application of model averaging to generalized functional linear models and proposed a leave-one-out cross-validation model averaging estimator as well. However, to our best knowledge, there is currently no literature investigating model averaging approaches specifically for partial linear functional additive models. The main purpose of this paper is to fill this gap. As FPC scores obtained from the functional predictor are employed for additive modeling, we need to deal with their empirical counterparts in regression models (see details in Section~\ref{sec2}). This introduces additional errors into the derivation of the Mallows-type criterion, rendering it no longer unbiased and potentially less efficient. Consequently, we adopt a more data-driven approach, namely multi-fold cross-validation model averaging. Moreover, compared to procedures such as jackknife or leave-one-out cross-validation, multi-fold cross-validation is computationally efficient. 

Our contributions can be summarized in three aspects. First, we introduce a novel model averaging framework specifically tailored to partial linear functional additive models. This framework effectively integrates multiple candidate models by assigning optimal weights based on a multi-fold cross-validation criterion. Second, we establish the asymptotic optimality of the model averaging estimator in terms of achieving the lowest square prediction error risk. We take into account the practical scenario where complete and precise records of the functional variable may not be available, and instead, we rely on discrete and finite observations with noise. Our investigation accounts for the subsequent challenge introduced by the estimated FPC scores and examines the optimality of the proposed approach in this context. This result provides a solid foundation for the applicability and reliability of our model averaging method when working with estimated FPC scores. Third, we provide empirical evaluations to support the proposed model averaging procedure. Extensive experiments and comparisons demonstrate the superiority of our model averaging method over conventional model selection techniques. By considering a range of candidate models and their corresponding weights, our approach can offer a more comprehensive and robust prediction. Overall, this paper contributes to the field of functional regression modeling by addressing model uncertainty through a model averaging approach.

The remainder of this paper is structured as follows. Section 2 introduces the model setup and presents the proposed model averaging estimator. The asymptotic optimality of the estimator is also established in Section 2. Section 3 reports the results of Monte Carlo simulations, while Section 4 illustrates an application to a NIR (Near Infra-red) spectra data. Section 5 concludes our work with a discussion. The proofs and additional simulation results can be found in the supplementary material.

\section{Methodology}\label{sec2}
\subsection{Model framework}\label{sec21}
The functional variable $U(t)$, $t\in\mathcal{T}$ is considered from the square integrable functional space $L^2(\mathcal{T})$, characterized by its mean function $\nu(t)$ and covariance function $\mathcal{C}(s,t)=cov\{U(s),U(t)\}$. Based on the eigen-decomposition of the corresponding covariance operator $(\mathcal{C}\psi_k)(t) = \lambda_k\psi_k(t)$, FPCA delivers a Karhunen-Lo\`{e}ve expansion of $U(t)$ as 
\[
U(t) = \nu(t) + \sum_{k=1}^{\infty} \zeta_k \psi_k(t), \: \zeta_k = \int_{\mathcal{T}} \{U(t)-\nu(t)\}\psi_k(t)dt, 
\]
where $\zeta_k$'s are uncorrelated variables known as functional principal component (FPC) scores. The variances of these FPC scores correspond to the eigenvalues $\{\lambda_k\}$, which satisfy $\lambda_1\geq \lambda_2\geq \cdots$. And $\{\psi_k\}$ is the sequence of eigenfunctions. Note that this representation of $U(t)$ is the most rapidly convergent one in the $L^2$ sense, making the leading FPC scores the most informative components capable of representing the information in $U(t)$. Building upon this insight, previous studies, such as \cite{yaoetal2005,caihall2006,hallhorowitz2007}, have employed FPC scores as predictors in regression, effectively transforming functional linear models into ordinary linear models. Similar techniques have also been applied to non-linear functional models, including functional additive models \citep{mulleryao2008,fanetal2015}, where FPC scores are used as predictors in an additive form. Inspired by these approaches, we adopt a same strategy in handling the functional predictor $U(t)$.

Suppose that $\big(Y_i, \bX_i, U_i(t)\big)$ are independent and identically distributed (i.i.d.) observations, where $t\in \mathcal{T}$ and $i=1,\ldots,n$. Here we consider the following partial linear functional additive model, which can be expressed as 
\begin{equation}\label{model:plfam}
	\begin{aligned}
		Y_i &= \mu_i + \ve_i = \bX_i^T\bbeta + \bff\big(\bxi_i\big) + \ve_i \\
		&= \sum_{j=1}^{p}X_{ij}\beta_j + \sum_{k=1}^{\infty}f_k\big(\xi_{ik}\big) + \ve_i, 
	\end{aligned}
\end{equation}
where the vector $\bX_i$ represents the scalar variables corresponding to the parametric components, and $\bbeta$ denotes the coefficient vector associated with these parametric components. The functions $f_k$ are assumed to be smooth functions. The vector $\bxi_i$ represents the transformed FPC scores, denoted as $\xi_{ik}$, where each score is obtained by applying $\Phi(\cdot)$ to the original FPC score, $\xi_{ik}= \Phi(\zeta_{ik};\lambda_k)$. The transformation function $\Phi$ maps values from $\mathbb{R}$ to $[0,1]$, which helps to avoid possible scale issues. 
The random error term $\ve_i$ is independent of both $\{\bX_i, U_i(\cdot)\}$. For ease of notation, we define $Y=(Y_1,\ldots,Y_n)^T$, $\mu=(\mu_1,\ldots,\mu_n)^T$, $\bX=(\bX_1,\ldots,\bX_n)^T$, and $\ve = (\ve_1,\ldots, \ve_n)^T$. Assume that the error term has a mean of $\bzero$ and a variance matrix denoted as $\bOmega=\mbox{diag}(\sigma_1^2, \cdots, \sigma_n^2)$. It is evident that when $p=0$, model (\ref{model:plfam}) simplifies to functional additive model proposed by \cite{mulleryao2008}. Here we approximate each additive component $f_k\big(\xi_{ik}\big)$ by spline basis functions $\{b_j(\cdot)\}_{j}$. This leads to the approximation $f_{k}\big(\xi_{ik}\big) \approx \bB_{ik}\big(\xi_{ik}\big)\bgamma_{k}$, where $\bB_{ik}\big(\xi_{ik}\big)$ denotes the vector of basis functions $\big\{b_j\big(\xi_{ik}\big)\big\}_{j}$. The corresponding basis matrix is denoted as $\bB_{k}\big(\xi_{k}\big) = (\bB_{1k}\big(\xi_{1k}\big), \bB_{2k}\big(\xi_{2k}\big), \ldots)^T$, and $\bgamma_{k}$ represents the coefficient vector associated with it. Therefore, 
\[
	\mu \approx \bX \bbeta + \sum_{k=1}^{\infty} \bB_{k}\big(\xi_{k}\big)\bgamma_{k} \equiv \bZ \btheta,
\]
where the design matrix is defined as $\bZ\equiv \big(\bX, \bB_{1}(\xi_{1}), \bB_{2}(\xi_{2}), \ldots\big)$, which comprises the design matrix $\bX$ and all the basis matrices $\bB_{k}(\xi_k)$. On the other hand, the coefficient vector for $\bZ$ is denoted as $\btheta\equiv \big(\bbeta, \bgamma_1, \bgamma_2, \ldots\big)^T$. 

Consider estimating $\btheta$ by minimizing penalized least squares objective function as follows, 
\[
\min_{\btheta} \|Y - \bZ\btheta\|^2 + \sum_{k=1}^{\infty} \tau_k\bgamma_k^T\bS_{k}\bgamma_k, 
\]
where $\|\cdot\|$ represents the $L_2$ norm of a vector, while $\bgamma_k^T\bS_{k}\bgamma_k$ is a smoothing penalty associated with the $k$-th additive component $f_k$. The smoothing matrix $\bS_{k}$ is constructed by considering the roughness of the basis functions, i.e., $\int b^{''}_{j_1}b^{''}_{j_2}$. The smoothing parameter $\tau_k$ controls the extent of penalization. As a result, the estimator of $\btheta$ can be expressed as 
\begin{equation}\label{wttheta}
	\wt{\btheta} = \Big(\bZ^T\bZ + \bS_{\tau}\Big)^{-1}\bZ^TY,
\end{equation}
where $\bS_{\tau}$ is a block diagonal matrix defined as $\bS_{\tau} = \mbox{diag}(\bzero_{p\times p}, \tau_1\bS_1, \tau_2\bS_2, \ldots)$. This implies that there is no regularization applied to the linear part, and only smoothing penalties are considered for the nonlinear components. 

However, it is noteworthy that having a perfectly and completely recorded functional variable $U(t)$ is not pragmatic in practice. Instead, we have access to a finite and discrete sampling of $U(t)$, which aligns with most data scenarios. Consequently, adjustments need to be made to the estimation procedure. Specifically, we consider that $U(t)$ is discretely recorded and contains additional measurement error, which can be expressed as
\[
U_{ij} = U_i(t_{ij}) + e_{ij}, \; i = 0, 1, \ldots, n, \; j=1, \ldots, N_i,
\]
where the measurement error $e_{ij}$ is assumed to be i.i.d. with mean $0$ and variance $\sigma_e^2$, and is independent of $U_i(t)$. We consider a dense and regular sampling design for $U(t)$, as it allows for efficient recovery of ${U_i(t)}$, see \cite{kongetal2016}. Denote the estimates of $\{\zeta_{ik}, \lambda_k, \psi_k\}_{i,k}$ based on the observed data $\{U_{ij}\}_{i,j}$ as $\{\wh{\zeta}_{ik}, \wh{\lambda}_k, \wh{\psi}_k\}_{i,k}$. The transformed FPC scores are then calculated as $\wh{\xi}_{ik} = \Phi\big(\wh{\lambda}_k^{-1/2}\wh{\zeta}_{ik}\big)$. The modified basis matrices, design matrix, and penalty matrix can be easily obtained as $\wh{\bB}_k\equiv \bB_k\big(\wh{\bxi}_k\big)$, $\wh{\bZ}=(\bX,\wh{\bB}_1,\wh{\bB}_2,\ldots)$, and $\wh{\bS}_{\tau}$, respectively. Furthermore, the penalized least squares criterion and the estimate for $\btheta$ are given by 
\[
	\min_{\btheta} \|Y-\wh{\bZ}\btheta\|^2 + \btheta^T\wh{\bS}_{\tau}\btheta
\]
and 
\[
	\wh{\btheta} = \big(\wh{\bZ}^T\wh{\bZ} + \wh{\bS}_{\tau}\big)^{-1}\wh{\bZ}^TY,
\]
respectively. Clearly, the performance of the model relies on the choice of regressors included, giving rise to model uncertainty.

\subsection{Model averaging by cross-validation criterion}\label{sec22}
We employ a set of $M$ candidate models to approximate the true model, where $M$ is a predetermined fixed number. In each candidate model, $p_m$ regressors in $\bX_i$ are included, while $q_m$ regressors in $\bxi_i$ are comprised. 
\[
Y_i = \mu_{(m),i} + \ve_{(m),i} = \bX_{(m),i}^T\bbeta_{(m)} + \bff_{(m)}\big(\bxi_{(m),i}\big) + \ve_{(m),i}, \quad m = 1, \ldots, M,
\]
where $\bX_{(m),i}$ represents the $i$-th row of $\bX_{(m)}$ with a corresponding $p_m$-dimensional coefficient vector $\bbeta_{(m)}$. $\bxi_{(m),i}$ is a $q_m$-dimensional vector, and $\ve_{(m),i}$ contains the approximation error of the $m$-th candidate model as well as the random error. Let $\wh{\bZ}_{(m)} = \big(\bX_{(m)}, \wh{\bB}_1, \ldots, \wh{\bB}_{q_m}\big)$ and 
$\wh{\bS}_{(m),\tau} = \mbox{diag}(0_{p_m\times p_m}, \tau_1\wh{\bS}_1, \ldots, \tau_{q_m}\wh{\bS}_{q_m})$. Hence, the estimator corresponding to the $m$-th candidate model can be expressed as
\begin{equation}\label{whtheta}
\wh{\btheta}_{(m)} = \big(\wh{\bZ}_{(m)}^T\wh{\bZ}_{(m)} + \wh{\bS}_{(m),\tau}\big)^{-1}\wh{\bZ}_{(m)}^TY,
\end{equation}
and therefore,
\[
\wh{\mu}_{(m),i} = \wh{\bZ}_{(m),i}^{T}\wh{\btheta}_{(m)}.
\]

The idea behind model averaging is to produce a final estimate by combining the estimates from all candidate models. To achieve it, we introduce a weight vector $\bomega=(\omega_1, \ldots, \omega_M)^T$ that belongs to the unit simplex of $\mathbb{R}^M$, denoted as $\mathcal{W} = \big\{\bomega\in [0,1]^M: \sum_{m=1}^M \omega_m = 1\big\}$. Consequently, the averaged estimate can be obtained as
\begin{equation}\label{whmu}
\wh{\mu}_i(\bomega) = \sum_{m=1}^{M}\omega_m\wh{\mu}_{(m),i} = \sum_{m=1}^{M}\omega_m\wh{\bZ}_{(m),i}^{T}\wh{\btheta}_{(m)}.
\end{equation}

The key question in model averaging estimation is how to combine estimates from candidate models by assigning weights. Our objective is to find a set of weights $\{\omega_m\}$ minimizes the expected squared prediction error loss for the resulting estimation $\wh{\mu}_{i}(\bomega)$. To formalize this, we define the prediction error risk as 
\[
r(\bomega) = \mathbb{E}\Big[\frac{1}{n}\sum_{i=1}^{n}\Big(Y_i^0 - \sum_{m=1}^{M}\omega_m\wh{\bZ}_{(m),i}^{T}\wh{\btheta}_{(m)}\Big)^2\Big], 
\]
where $Y_i^0$ is generated from model (\ref{model:plfam}) with a seperate random error $\ve_i^0$ replacing $\ve_i$. This implies that $Y_i^0$ is an independent observation from the same model. The $r(\bomega)$ above resembles the squared error risk used in previous studies such as \cite{hansen2007}, \cite{wanetal2010}, and \cite{baietal2022}. 

Specifically, the data are divided into Q disjoint subsamples of approximately equal size $L = n/Q$, with each subsample serving as a validation set in a sequential manner. The detailed procedure is outlined below. Note that when $Q=n$, the Q-fold cross-validation reduces to leave-one-out cross-validation, which can be computationally demanding for large sample sizes.

\begin{enumerate}[\it{Step} 1.]
	\item Estimate the mean function $\wh{\nu}(t)$, the covariance function $\wh{\mathcal{C}}(s,t)$, and the transformed FPC scores $\{\wh{\xi}_{ij}\}$ for the functional variable $U(t)$.
	
	\item For each candidate model, exclude the $q$-th subsample and obtain the estimate $\wh{\btheta}_{(m)}^{[-q]}$ using the remaining $Q-1$ subsamples, where $q=1,\ldots,Q$. 
	
	\item Predict the response values of the $q$-th subsample using $\wh{\btheta}_{(m)}^{[-q]}$, i.e., 
	\begin{equation}\label{yqfold}
	\wh{Y}^{[-q]}_{(m),(q-1)L+l} = \wh{\bZ}_{(m),(q-1)L+l}^{T}\wh{\btheta}^{[-q]}_{(m)}, \; l=1,\ldots,L.
	\end{equation}
	
	\item Combine the predictions from all candidate models for the $q$-th subsample by averaging, resulting in 
	\[
	\wh{Y}^{[-q]}_{(q-1)L+l}(\bomega) = \sum_{m=1}^M \omega_m \wh{Y}^{[-q]}_{(m),(q-1)L+l}.
	\]
\end{enumerate}

The Q-fold cross-validation criterion is then defined as 
\begin{equation}\label{cvqomega}
	CV_Q(\bomega) = \sum_{q=1}^{Q}\sum_{l=1}^{L} \Big(Y_{(q-1)L+l} - \sum_{m=1}^{M}\omega_{m}\wh{\bZ}^T_{(m),(q-1)L+l}\wh{\btheta}_{(m)}^{[-q]}\Big)^2 ,
\end{equation}
and the weight vector for averaging is determined by 
\begin{equation}\label{quadprog}
	\wh{\bomega} = \arg\min_{\bomega\in\mathcal{W}} CV_Q(\bomega).
\end{equation}
Once $\wh{\bomega}$ is obtained, we could construct a cross-validation model averaging estimation $\wh{\mu}_i(\wh{\bomega})$.

\subsection{Computation}\label{sec23}
To efficiently compute the weights $\wh{\bomega}$, we reformulate the cross-validation criterion (\ref{cvqomega}-\ref{quadprog}) as a quadratic programming problem. Let $\wh{\bY}^{[-q]}$ be an $n\times M$ matrix with elements $\big\{\wh{Y}^{[-q]}_{(m),(q-1)L+l}\big\}$, and $\bY = (Y,\ldots, Y)$ be an $n\times M$ matrix. Then, we have 
\begin{equation}\label{quadprog2}
\begin{aligned}
	CV_Q(\bomega) &= \sum_{q=1}^{Q} \sum_{l=1}^{L} \Big(Y_{(q-1)L+l} - \sum_{m=1}^M \omega_m \wh{Y}_{(m),(q-1)L+l}^{[-q]} \Big)^2 \\
	&= \big( Y - \wh{\bY}^{[-q]}\bomega \big)^T \big( Y - \wh{\bY}^{[-q]}\bomega \big) \\
	&= \bomega^T \big( \bY - \wh{\bY}^{[-q]} \big)^T \big( \bY - \wh{\bY}^{[-q]} \big) \bomega.
\end{aligned}
\end{equation}
Hence, the optimization problem (\ref{quadprog}) can be formulated as a standard quadratic programming problem (\ref{quadprog2}) with the constraints $\{\omega_m \geq 0\}$ and $\sum_{m=1}^M \omega_m = 1$. The computation of the proposed procedure is formally outlined in Algorithm~\ref{algo1}.

\begin{algorithm}
\caption{Estimate $\wh{\bomega}$ and predict $\wh{\mu}_i(\wh{\bomega})$}\label{algo1}
\begin{algorithmic}[1]
\Require Data $(Y_i, \bX_i, U_i)$, $M$ candidate models, and number of folds $Q$. 
\Ensure $\wh{\bomega}$ and $\wh{\mu}_i(\wh{\bomega})$.
\State Apply FPCA on $\{U_i(t)\}$ to obtain the covariance function $\mathcal{C}(s,t)$, take eigen-decomposition and get the estimated transformed FPC scores $\{\wh{\xi}_{ij}\}$, generate basis matrices $\wh{\bB}_k=\bB_k\big(\wh{\bxi}_k\big)$.
\While{$m$ in $1:M$}
    \State Prepare data $(\bY, \wh{\bZ}_{(m)})=(\bY, \big(\bX_{(m)}, \wh{\bB}_1, \ldots, \wh{\bB}_{q_m}\big))$ according to the $m$-th candidate model. 
    \State Randomly partition the data into $Q$ folds.
    \While{$q$ in $1:Q$}
        \State Exclude the $q$-th subsample and obtain the estimate $\wh{\btheta}_{(m)}^{[-q]}$ by Equation (\ref{whtheta}) using the remaining $Q-1$ subsamples. 
        \State Predict the response values of the $q$-th subsample, $\wh{Y}^{[-q]}_{(m),(q-1)L+l}$, by Equation (\ref{yqfold}), $l=1,\ldots,L$.
    \EndWhile
\EndWhile
\State Estimate $\wh{\bomega}$ by solving the quadratic programming problem (\ref{quadprog2}) with constraints $\bomega\geq 0$ and $\bomega^T\mathbf{1}=1$.
\State Obtain the final averaging prediction $\wh{\mu}_i(\wh{\bomega})$ by substituting $\wh{\bomega}$ into Equation (\ref{whmu}).
\end{algorithmic}
\end{algorithm}

Algorithm~\ref{algo1} provides a systematic approach to estimate the optimal weights for combining models through a Q-fold cross-validation criterion. Each candidate model is trained and evaluated on different folds of the data. We repeat the process for each candidate model, and update the weights based on the prediction errors on the validation folds. Finally, the optimal weights and the corresponding prediction are outputted.

\begin{remark}
The candidate models constructed above only differ in the included regressors. However, it is possible to create additional candidate models by varying tuning parameters, such as the number of interior knots in the basis functions, denoted as $J_n$, or the smoothing parameter $\tau$. It is noteworthy that incorporating information from a large number of candidate models can potentially lead to higher prediction accuracy. However, this comes at the cost of increased computational resources required for model estimation and evaluation. 
\end{remark}

\subsection{Asymptotic optimality}\label{sec3}
Before establishing the theoretical results, we introduce relevant notations and assumptions. Let $\bar{p} = \max_{1\leq m\leq M} p_m$ and $\bar{q} = \max_{1\leq m\leq M} q_m$. Assume that for each candidate model, the estimator $\wt{\btheta}_{(m)}$ in (\ref{wttheta}) converges to a limit $\btheta_{(m)}^{*}$ satisfying that $\|\btheta_{(m)}^{*}\|^2/(\bar{p} + \bar{q}J_n)\leq C_{\theta}<\infty$, where $C_{\theta}$ is a constant. 

Let 
\[
\overline{R}(\bomega) = \mathbb{E}\Big[\frac{1}{n}\sum_{i=1}^n \Big(\mu_i - \sum_{m=1}^{M} \omega_m\wh{\bZ}^T_{(m),i}\wh{\btheta}_{(m)}\Big)^2\Big],
\]
and 
\[
R^{*}(\bomega) = \frac{1}{n}\sum_{i=1}^{n} \Big(\mu_i - \sum_{m=1}^{M}\omega_{m}\bZ_{(m),i}^{T}\btheta_{(m)}^{*}\Big)^2.
\]

Denote $\mathcal{S}_n$ as the space of polynomial splines on the interval $[0,1]$ with degree $\rho$. When $\rho\geq 2$, each spline function in $\mathcal{S}_n$ is $\rho-1$ continuously differentiable on $[0,1]$. Let $\eta_n = \inf_{\bomega\in \mathcal{W}} nR^*(\bomega)$. In the current development, our focus is on fixed values of $p_m$ and $q_m$. The following conditions are required for the theoretical derivation. 

\begin{condition}\label{fdcondition}
\begin{enumerate}[(a)]
    \item The eigenvalues of the functional variable $U(t)$ satisfy 
          \[
           c_{\lambda}^{-1}k^{-\alpha}\leq \lambda_k \leq C_{\lambda}k^{-\alpha}, \quad \lambda_k - \lambda_{k+1} \geq C_{\lambda}^{-1}k^{-1-\alpha}, k=1,2,\ldots,
           \]
           where $c_{\lambda}$ and $C_{\lambda}$ are nonzero constants. Furthermore, we assume that $\alpha>1$ to ensure that $\sum_{k=1}^{\infty} \lambda_k < \infty$. 
    \item Assume that $\mathbb{E}(\|U\|_{\mathbb{U}}^4)<\infty$, where $\|\cdot\|_{\mathbb{U}}$ represents the norm in $L^2(\mathcal{T})$, defined as $\|U\|_{\mathbb{U}} = \big(\int_{\mathcal{T}}U^2(t)dt\big)^{1/2}$. Additionally, there exists a constant $C_{\zeta}>0$ such that $\mathbb{E}(\zeta_k^2\zeta_j^2)\leq C_{\zeta}\lambda_k\lambda_j$ and $\mathbb{E}(\zeta_k^2-\lambda_k)^2<C_{\zeta}\lambda_k^2$, $\forall k, k\neq j$.
\end{enumerate}
\end{condition}

Condition \ref{fdcondition} mainly restricts the behaviour of the functional predictor $U(t)$. Condition \ref{fdcondition}(a) assumes that the eigenvalues decay at a polynomial rate, which is a relatively slow rate and allows for flexibly modeling of $U(t)$ as an $L^2(\mathcal{T})$ process. Additionally, it requires that eigenvalue spacings are not too small, guaranteeing the identifiability and consistency of sample eigenvalues and eigenfunctions. Condition \ref{fdcondition}(a) is commonly used in the literature of functional data modeling, see \cite{caihall2006,caiyuan2012}. Condition \ref{fdcondition}(b) is a weak moment restriction on $U(t)$. It is satisfied when $U(t)$ follows a Gaussian process, as discussed in \cite{wongetal2019}. Given that the FPC scores are employed to model the effect of $U(t)$ in this context, it is reasonable to ensure the effectiveness of the estimated FPC scores by Condition \ref{fdcondition}(b).


\begin{condition}\label{splinecondition}
\begin{enumerate}[(a)]
    \item The functions $\{f_k\}$ in model (\ref{model:plfam}) belong to a class of functions $\mathcal{H}$ defined on the interval $[0,1]$. These functions have $\iota$-th derivatives that exist and satisfy a Lipschitz condition of order $\gamma$. Specially, for any $z', z \in [0,1]$, we have 
    \[
    \big| g^{(\iota)}(z')-g^{(\iota)}(z) \big|\leq C_g\big| z'-z \big|^{\gamma},
    \]
    where $C_g$ is a positive constant. Here, $\iota$ is a positive integer and $\gamma\in(0,1]$ such that $s =\iota +\gamma >1.5$.
          
    \item The number of interior knots for spline approximation, denoted by $J_n$, satisfies that $n^{1/2s}\ll J_n\ll n^{1/3}$. When $s = 2$, a suitable choice for $J_n$ is $n^{1/4}\log n$. 
\end{enumerate}
\end{condition}

Condition \ref{splinecondition} is a widely-used assumption in nonparametric modeling when employing spline approximation for smoothing functions, see \cite{stone1982,fanetal2011,liuetal2011} for example. 

\begin{condition}\label{momentcondition}
\begin{enumerate}[(a)]
    \item $\sup_{i=1,\ldots,n}\mathbb{E}\ve_i^2\leq C_{\ve}<\infty$, where $C_{\ve}$ is a constant.
    \item $\mu_j = O(1)$, a.s., $j=1,\ldots,n$.
    \item $\sup_{j=1,\ldots,n} \|\bZ_{(m),j}\|^2/(\bar{p}+\bar{q}J_n)\leq C_z<\infty$, where $C_z$ is a constant.
\end{enumerate}
\end{condition}

Condition \ref{momentcondition}(a) ensures that the random error has finite variance, which is necessary for asymptotic analysis of nonlinear least square estimation, refer to \cite{wu1981}. Condition \ref{momentcondition}(b) is equivalent to $\|\mu\|^2/n = O(1)$. It is commonly used in regression problems and helps exclude inflation cases in the limiting process. Condition \ref{momentcondition}(c) is satisfied when $\|\bX_{(m),j}\|^2/p_m$ and $\{\|\bB_{(m),j}\|^2/J_n\}$ are finite. Examples of such cases include the normalized version of the scalar regressors $\bX_{(m),j}$ and the spline basis $b_j(\cdot)$. Similar conditions can be found in other works that deal with partial linear models, see \cite{liuetal2011} for example. The normalization is not essential but is imposed to simplify certain expressions in our theoretical development.

\begin{condition}\label{ordercondition}
$n^{1/2}J_n\eta_n^{-1} = o_p(1)$, $nh^{s}J_n\eta_n^{-1}=o_p(1)$, and $J_n^{2}\eta_n^{-1}=o_p(1)$.
\end{condition}

Condition \ref{ordercondition} imposes restrictions on the growth rate of $\eta_n$. Similar conditions can be found in other works, such as condition (21) in \cite{zhangetal2013}, condition (7) in \cite{andoli2014}, and Condition 3 in \cite{zhangetal2018}. These technical conditions place constraints on the growth rate of $\eta_n$ to ensure optimality. For instance, when $s=2$ and $J_n=n^{1/4}\log n$ is taken, Condition \ref{ordercondition} can be simplified to $n^{3/4}(\log n)\eta_n^{-1} = o_p(1)$ and $n^{1/2}(\log n)^2\eta_n^{-1} = o_p(1)$. This implies that $\eta_n$ grows at a rate larger than $n^{1/2}$. Consequently, Condition \ref{ordercondition} requires that all candidate models have nonzero bias, indicating that they are misspecified.

\begin{theorem}\label{thm1}
	Suppose Conditions \ref{fdcondition}--\ref{ordercondition} hold. As the sample size $n$ tends to $\infty$, we have 
	\begin{equation}\label{finalgoal}
		r(\wh{\bomega}) / \inf_{\bomega \in \mathcal{W}} r(\bomega) \to 1
	\end{equation}
	in probability.
\end{theorem}

Theorem~\ref{thm1} means that the ratio of the risk achieved by the estimated weights $\wh{\bomega}$ to the infimum risk over the weight space $\mathcal{W}$ converges to 1 in probability. It illustrates the asymptotic optimality of $\wh{\bomega}$, as it yields a prediction risk that is asymptotically identical to that of the infeasible optimal weight vector. Therefore, the resulting cross-validation model averaging estimator also enjoys the asymptotic optimality in terms of square error loss. 


\section{Simulation studies}\label{sec4}
This section aims to evaluate the finite sample performance of the proposed Cross-Validation Model Averaging (CVMA) estimator. We present the results of both 5-fold and 10-fold CVMA to illustrate its efficacy. To demonstrate the superiority of the CVMA estimator over other information criteria-based methods, we compare it with the AIC model selection (AIC), BIC model selection (BIC), smoothed AIC model averaging (SAIC), and smoothed BIC model averaging (SBIC). AIC and BIC choose a single model based on the lowest score defined as 
\[
\begin{aligned}
	AIC_m &= n\log(\wh{\sigma}_m^2) + 2df_m, \\
	BIC_m &= n\log(\wh{\sigma}_m^2) + \log(n)df_m,
\end{aligned}
\]
while SAIC and SBIC allocate weights to each candidate model based on their respective AIC and BIC scores
\[
\omega_m^{AIC} = \frac{\exp(-AIC_m/2)}{\sum_{m=1}^M \exp(-AIC_m/2)}
\]
and 
\[
\omega_m^{BIC} = \frac{\exp(-BIC_m/2)}{\sum_{m=1}^M \exp(-BIC_m/2)},
\]
where $\wh{\sigma}_m^2 = \|Y-\wh{Y}_{(m)}\|^2/n$ and $df_m$ represents the effective degree of freedom of the $m$-th candidate model, see \citet{akaike1973,schwarz1978,bucklandetal1997}.

The data generating process is
\[
Y_i = \mu_i + \varepsilon_i = \sum_{j=1}^{M_0}\beta_j X_{ij} + \sum_{k=1}^{K_0} f_k(\xi_{ik}) + \ve_i, \quad i = 1, \ldots, n,
\]
where $\xi_{ik}$'s are the transformed FPC scores from the original $\zeta_{ik}$'s. $\zeta_{ik}$, $i=1,\ldots,n$ are drawn independently from $N(0, \lambda_k)$, and take $\Phi(\cdot)$ as the standard Gaussian cumulative distribution function. The test data $Y_i^0$'s generate from $Y_i^0 = \mu_i + \ve_i^0$, where $\ve_i^0$'s are independent from $\ve_i$'s. We consider the following setups.

\noindent\textbf{Design 1.} $M_0 = 50$ and $\beta_j = j^{-3/2}$. $\bX_i$'s are i.i.d. observations from $MN(0, \Sigma_{M_0})$, where the $a,b$-th element $\Sigma_{ab}$ of $\Sigma_{M_0}$ equals $0.5^{|a-b|}$. We set $K_0 = 50$ and generate the functional variable $U_i(t)$ as follows: 
\[
U_i(t) = \sum_{k=1}^{50} \zeta_{ik}\psi_k(t),\quad t\in\mathcal{T} = [0,1],
\]
where $\zeta_{ik}\sim N(0, k^{-3/2})$, $\psi_k(t) = \sqrt{2}\sin(k\pi t)$, $k = 1,\ldots,50$. The error terms $\varepsilon_i$ are homoscedastic with $\varepsilon_i \sim N(0, \eta^2)$. We vary $\eta$ so that $R^2 = \text{var}(\mu_i) / \text{var}(Y_i)$ ranges from 0.1 to 0.9, where $\text{var}(\mu_i)$ and $\text{var}(Y_i)$ represent the variances of $\mu_i$ and $Y_i$, respectively. The additive effect of $U(t)$ is described as 
\[
\sum_{k=1}^{50} f_k(\xi_k) = \frac{3}{2}\Big\{(\xi_1-\frac{1}{2})^2-\frac{1}{12}\Big\} + (\xi_2-\frac{1}{2}) + \frac{3}{2}(\sin\pi\xi_3 -\frac{2}{\pi}) + \sum_{k=4}^{50}\frac{1}{k}(\xi_k-\frac{1}{2}).
\]
It is worth noting that each additive component's expectation is standardized to 0 for the purpose of identifiability. Additionally, the scalar variables $\bX$ and the functional variable $U(t)$ in Design 1 are independent of each other.

\noindent\textbf{Design 2.} $M_0 = 50$ and $\beta_j = j^{-1/2}$. Consider the scenario that $\bX$ and $U(t)$ are correlated, which can be simulated by drawing $(\bX_i, \zeta_{i1})\sim MN(0, \Sigma_{M_0+1})$, where $\Sigma_{ab}$ denotes the $a,b$-th element of the covariance matrix $\Sigma_{M_0+1}$, which is defined as $\Sigma_{ab}=0.5^{|a-b|}$. It is clear that the correlation between $\bX$ and $U(t)$ is induced through the correlation structure of the elements in $\Sigma_{M_0+1}$. $K_0 = 20$ and the functional variable $U_i(t)$ is generated by
\[
U_i(t) = \sum_{k=1}^{20} \zeta_{ik}\psi_k(t),\quad t\in\mathcal{T} = [0,10],
\]
where $\zeta_{ik}\sim N(0, k^{-2})$, $k=2,\ldots,20$. $\psi_k(t) = \cos(k\pi t/5)/\sqrt{5}$, $k = 1,\ldots,20$. The error terms $\ve_i$ are heteroscedastic as $\ve_i\sim N\big(0, \eta^2(u_{i}^2+0.01)\big)$, where $u_i$ is a uniform random variable drawn from the interval $[-1,1]$. Still vary $\eta$ so that the $R^2$ values fall within the range of 0.1 to 0.9. And the additive effect is 
\[
\sum_{k=1}^{20} f_k(\xi_k) = 2(\xi_1-\frac{1}{2}) + \frac{3}{2}(\exp(\xi_2)-e+1) + (\xi_3-\frac{1}{2})^2-\frac{1}{12} + \sum_{k=4}^{20}\frac{3}{k}(\xi_k-\frac{1}{2}).
\]
To ensure identifiability, the expectation of each additive component is also centered at 0 by standardization.

\noindent\textbf{Design 3.} $M_0 = 50$ and $\beta_j = j^{-1}$. We adopt the same approach as in Design 2 to simulate correlated $\bX$ and $U(t)$ through $(\bX_i, \zeta_{i1})\sim MN(0, \Sigma_{M_0+1})$, where the $a,b$-th element of $\Sigma_{M_0+1}$ is $\Sigma_{ab}=0.5^{|a-b|}$. $K_0=50$ and $U_i(t)$ is constructed by 
\[
U_i(t) = \sum_{k=1}^{50} \zeta_{ik}\psi_k(t),\quad t\in \mathcal{T} = [0,1],
\]
where $\zeta_{ik}\sim N(0, k^{-3/2})$, $k=2,\ldots,50$. $\psi_k(t) = \sqrt{2}\sin(k\pi t)$, $k = 1,\ldots,50$. The error terms $\ve_i$ are heteroscedastic as $\ve_i\sim N\big(0, \eta^2(X_{i2}^2+0.01)\big)$. Vary $\eta$ so that $R^2 = var(\mu_i)/ var(Y_i)$ values range from 0.1 to 0.9. The additive function is the same as that of Design 1. To summarize, Design 3 shares similarities with Design 1 in terms of the setup of $U(t)$ and $\bX$, but differs in the presence of correlated regressors, coefficients $\beta_j$, and heteroscedastic random errors.

\begin{figure*}[!htb]
	\centering
	\includegraphics[width=\textwidth]{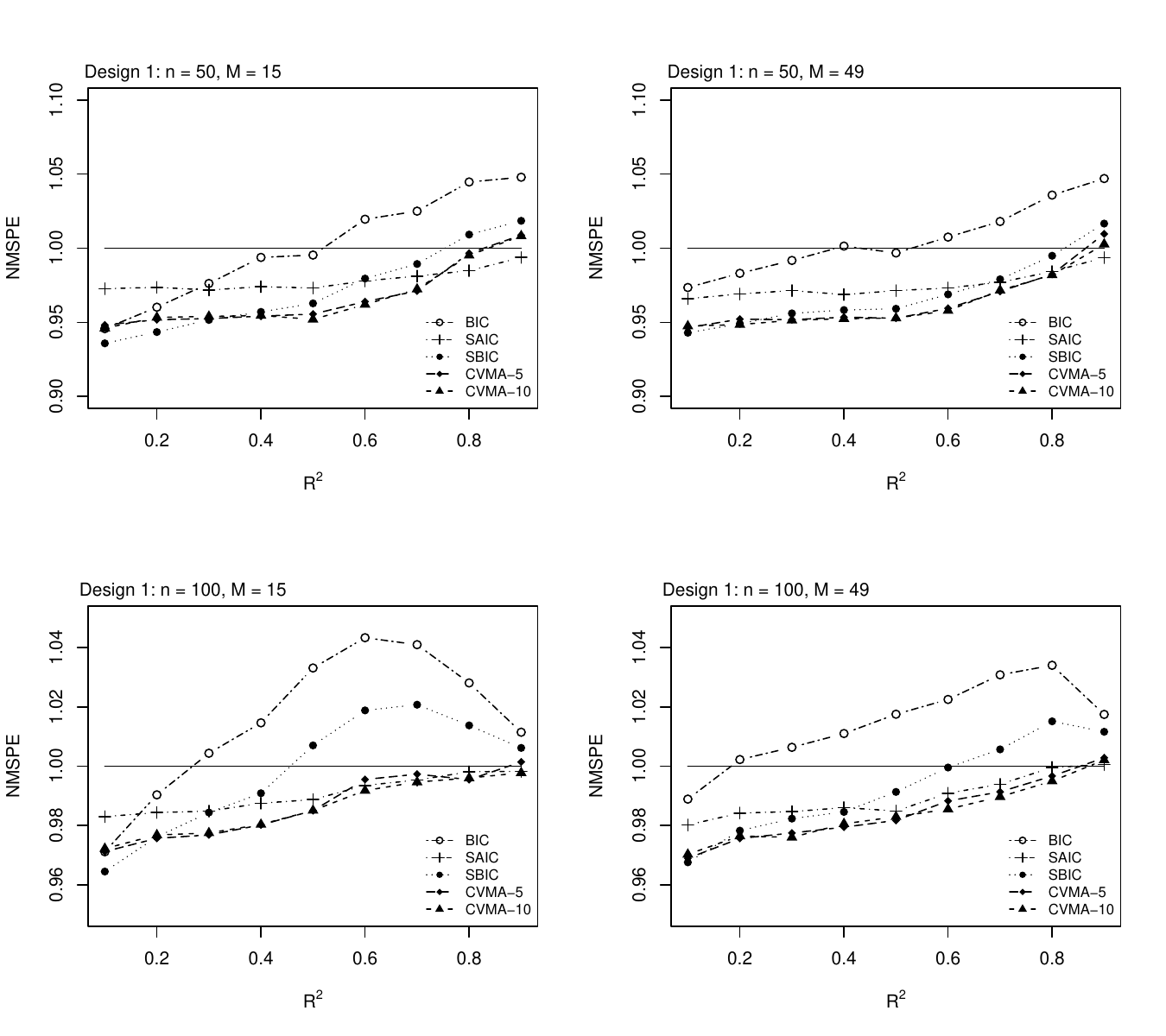}
	\caption{Simulation results of NMSPE values for Design 1 over $D=200$ simulations.}
	\label{figd1}
\end{figure*}

In each of the designs, the functional predictor $U(t)$ is observed at 100 equally-spaced grids on $\mathcal{T}$ with i.i.d. measurement errors $e_{ij}$'s that follow $N(0,0.2)$ distribution. Specifically, the $i$-th observation of $U(t)$ at $t_j$ is denoted by $U_{ij} = U_i(t_j) + e_{ij}$. The training set sample size is fixed at $n=50$ or $n=100$, and the test set sample size $n_{test}$ is set to be the same as the training set. To approximate $f_k$'s, B-spline bases of order 4 are employed, and the smoothing parameters are set as $\tau_0 \equiv \tau_1=\tau_2=\cdots$ for simplicity, where $\tau_0$ can be selected by the maximum likelihood method proposed by \cite{wood2017}. Consider two kinds of settings for candidate models:

\begin{enumerate}
	\item The nested setting: each candidate model consists of the first several components of both $\bX$ and $\bxi$. Specifically, the set of candidate scalar variables is $\{X_1,\ldots,X_5\}$, while the set of candidate transformed FPC scores is comprised of $\{\xi_1,\xi_2,\xi_3\}$. Each candidate model includes at least one scalar variable and one transformed FPC score, leading to a total of $M = 5\cdot 3 = 15$ candidate models. The nested structure is often employed when the analyst has some background or prior knowledge about the structure of models. 
	
	\item The non-nested setting: assume that at least one scalar variable in $\{X_1,X_2,X_3\}$ and one transformed FPC score in $\{\xi_1,\xi_2,\xi_3\}$ are included in a candidate model. As a result, we have a total of $M = \left[\binom{3}{3}+\binom{3}{2}+\binom{3}{1}\right]\cdot \left[\binom{3}{3}+\binom{3}{2}+\binom{3}{1}\right] = 49$ candidate models. This setting is particularly suitable when there is no prior knowledge or assumptions about the structure of the models.
	
\end{enumerate}

\begin{figure*}[!htb]
	\centering
	\includegraphics[width=\textwidth]{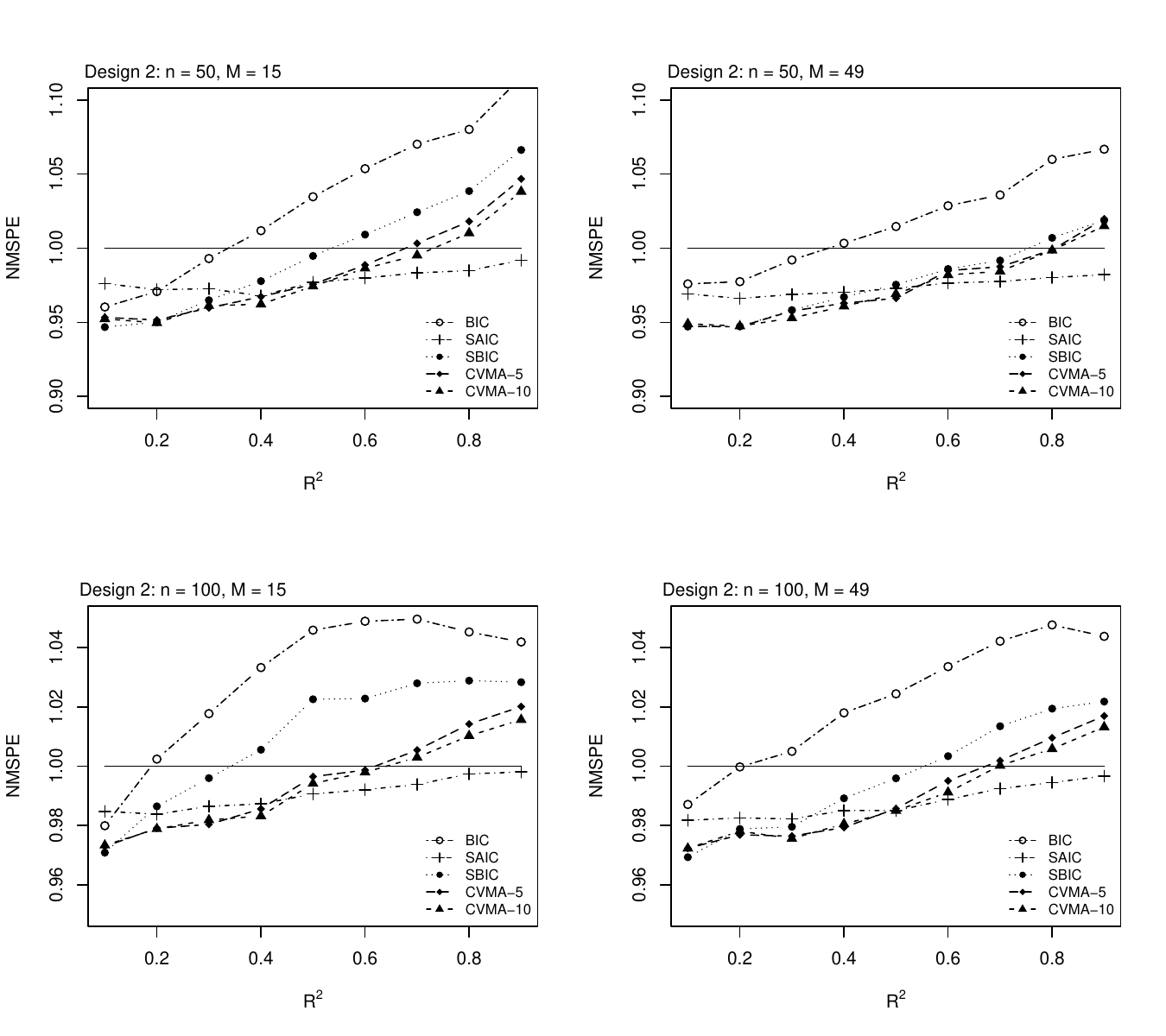}
	\caption{Simulation results of NMSPE values for Design 2 over $D=200$ simulations.}
	\label{figd2}
\end{figure*}


We assess the performance of all methods based on the averaged mean square prediction error (MSPE) over $D=200$ replications on the test set 
\[
MSPE = \frac{1}{Dn_{test}}\sum_{d=1}^{D}\big\|Y^0-\wh{\mu}(\wh{\bomega})\big\|^2,
\]
where $d$ represents the $d$-th trial. To facilitate comparisons, we normalize all MSPE values by dividing them with the MSPE value of the AIC estimator. Consequently, a normalized MSPE (NMSPE) less than 1 indicates that the corresponding estimator outperforms the AIC estimator, and vice versa. The comparison results of the standard errors of root-MSPE values are presented in the supplementary material.

\begin{figure}[!htb]
	\centering
	\includegraphics[width=\textwidth]{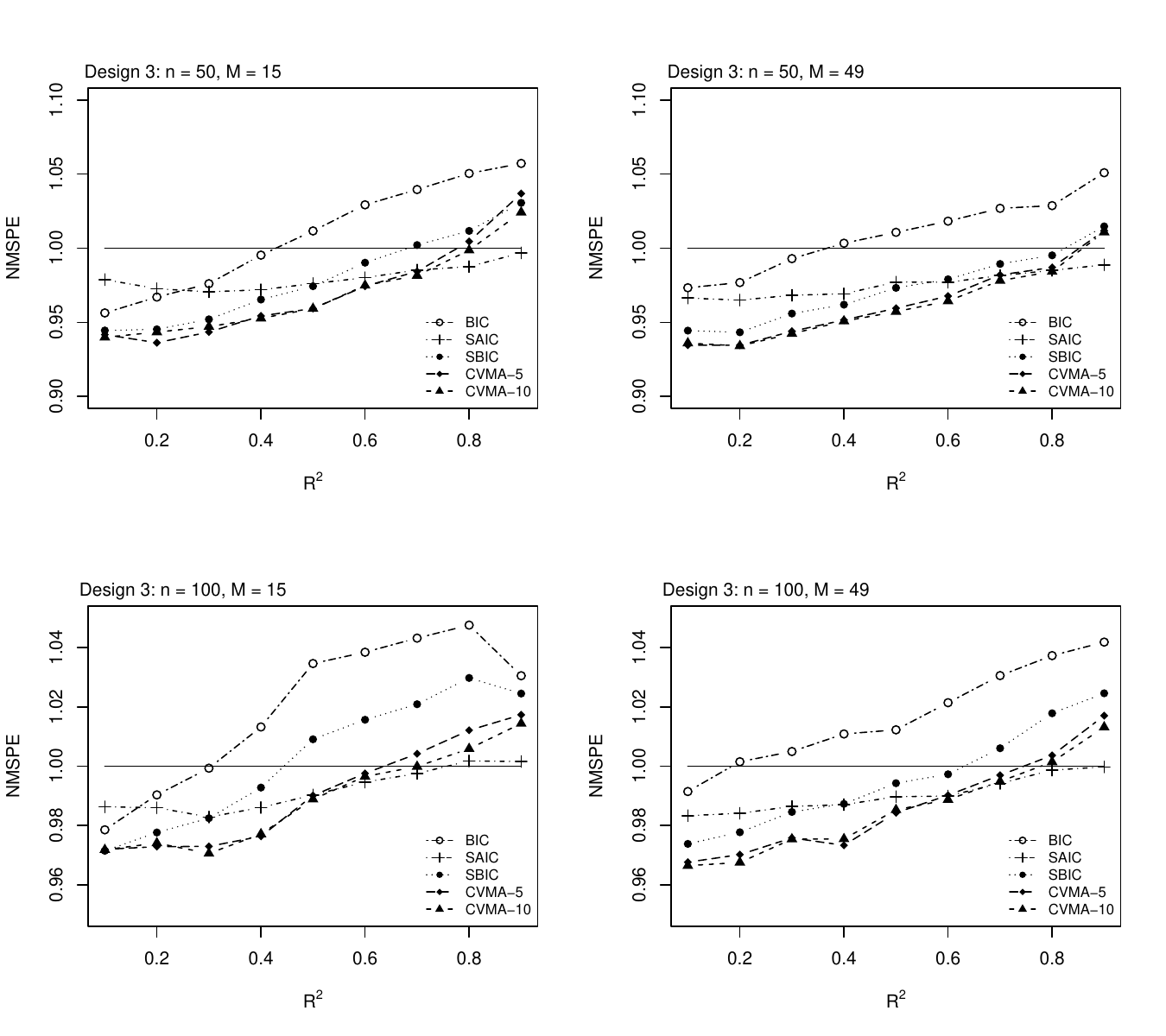}
	\caption{Simulation results of NMSPE values for Design 3 over $D=200$ simulations.}
	\label{figd3}
\end{figure}

The simulation results of NMSPEs for each design are displayed in Figures~\ref{figd1}-\ref{figd3}. 
Figure~\ref{figd1} illustrates that the 5-fold CVMA (CVMA-5) and 10-fold CVMA (CVMA-10) estimators have comparable performance, where the 10-fold CVMA performs slightly better than the 5-fold CVMA, and both methods generally outperform the other procedures. For small $R^2$ values, BIC, SAIC, SBIC, and CVMA (CVMA-5, CVMA-10) methods produce smaller NMSPE values than AIC. However, BIC and SBIC, particularly BIC, show significant deterioration as the $R^2$ value increases or the sample size $n$ grows, while SAIC and CVMA maintain their advantages. This suggests that compared with other methods, BIC selection method loses its edge in prediction when more information (e.g., stronger signals or larger training sample size) is included in the training process, which is consistent with the parsimonious nature of BIC. On the other hand, SBIC, as an averaging method, consistently performs better than BIC in prediction but also deteriorates under such scenarios due to its BIC-based nature. Furthermore, the NMSPE values of SAIC are smaller than or close to that of AIC for all $R^2$ values, indicating that SAIC can improve prediction accuracy to some extent. And for large $R^2$ values, although CVMA can lead to larger MSPE values than AIC and SAIC, the difference narrows as the training sample size or the number of useful candidate models increases. From this perspective, the CVMA estimator is either superior or comparable to the other methods. Overall, the simulation results for Design 1 suggest that the 5-fold CVMA and the 10-fold CVMA estimators are promising alternatives to other model selection estimators.

Figures~\ref{figd2} and \ref{figd3} show the NMSPE results for more complicated cases, in which the regressors are correlated, and the random errors are heteroscedastic. In Figure~\ref{figd2}, it can be observed that CVMA performs the best for small and medium $R^2$ values, while it performs worse than AIC and SAIC for large $R^2$ values. Furthermore, for medium and large $R^2$ values in Design 2, 10-fold CVMA exhibits a clear advantage over 5-fold CVMA, indicating the benefits of effective information in the training process. Moreover, BIC performs better than AIC and SAIC only for small $R^2$ values and deteriorates significantly as $R^2$ increases or $n$ grows, which is similar to its performance in Design 1. Although SBIC shows a similar trend to CVMA, it delivers a larger MSPE value than that of CVMA. In Figure~\ref{figd3}, similar changes to those in Figure~\ref{figd2} are presented, and hence similar conclusions can be drawn.

In conclusion, model averaging procedures generally outperform model selection procedures in terms of predictive performance in most scenarios considered in this study. Model selection procedures rely solely on one model and may miss the benefits of other useful candidate models. The CVMA estimator yields satisfactory prediction outcomes in most cases, including situations with small and medium $R^2$ values or limited training sample size. This finding indicates that CVMA works better than SAIC when the available information is limited, demonstrating the efficiency and effectiveness of the CVMA estimator. However, the presence of correlated regressors and heteroscedastic error terms can diminish the comparative advantage of CVMA in terms of prediction accuracy. Additionally, SBIC performs well for small $R^2$ values, while SAIC performs better for larger $R^2$ values. This finding implies that the advantage of SAIC increases as useful information increases in the training process. Finally, the NMSPE values show slightly different trends for $M=15$ and $M=49$, suggesting that the construction of useful candidate models should also be considered.

\section{Real data analysis}\label{sec5}
In this section, we employ the proposed CVMA method to analyze the {\it NIR shootout 2002} dataset, which was published by the International Diffuse Reflectance Conference (IDRC) in 2002 and is available from Eigenvetor Research Inc, USA.\footnote{http://www.eigenvector.com/data/tablets/index.html} The dataset comprises of NIR (Near Infra-red) spectra (functional variable $U(t)$) of 655 pharmaceutical tablets, which are measured using two spectrometers over the spectral region ranging from 600 to 1898 nm with 2 nm increments on the wavelength scale. Additionally, quantities such as the weight of active ingredient (response variable $Y$), tablet hardness (scalar predictor $X_1$), and tablet weight (scalar predictor $X_2$) are also provided for reference analysis. The data have been pre-divided into training (155 tablets), validation (40 tablets) and test (460 tablets) subsets. We used the spectra records from instrument 1 for analysis, and pre-standardized the sample data of $Y$, $X_1$, and $X_2$. Then, candidate models were trained on the training subset (TRAIN), and their performances were evaluated on the test subset (TEST). Finally, mean square prediction error (MSPE) was used to compare the predictive efficiency,
\[
MSPE = \frac{1}{n_{test}}\sum_{i\in \mbox{\tiny TEST}}\big(Y_{i} - \wh{\mu}_{i}\big)^2,
\]
where $n_{test}$ is the size of TEST. In this study, we constructed candidate models using both parametric (scalar predictors $X_1$ and $X_2$ ) and nonparametric (transformed FPC scores $\{\xi_1,\xi_2,\ldots,\xi_7\}$, accounting for at least 99.5\% of variance explained in $U(t)$) components. The non-nested structure was adopted due to the lack of prior knowledge about the model structure. Therefore, each candidate model consisted of at least one component from each part, resulting in a total of $M = 87$ candidate models. We compared the MSPE of CVMA with that of AIC, BIC, SAIC, SBIC, and equally-weighting (simple averaging) methods. The cross-validation procedure involved randomness from training data splitting, so CVMA was replicated 100 times for illustration, the other methods utilized all the training data and provided a single MSPE value.

\begin{figure}[!htb]
	\centering
	\includegraphics[width=0.75\textwidth]{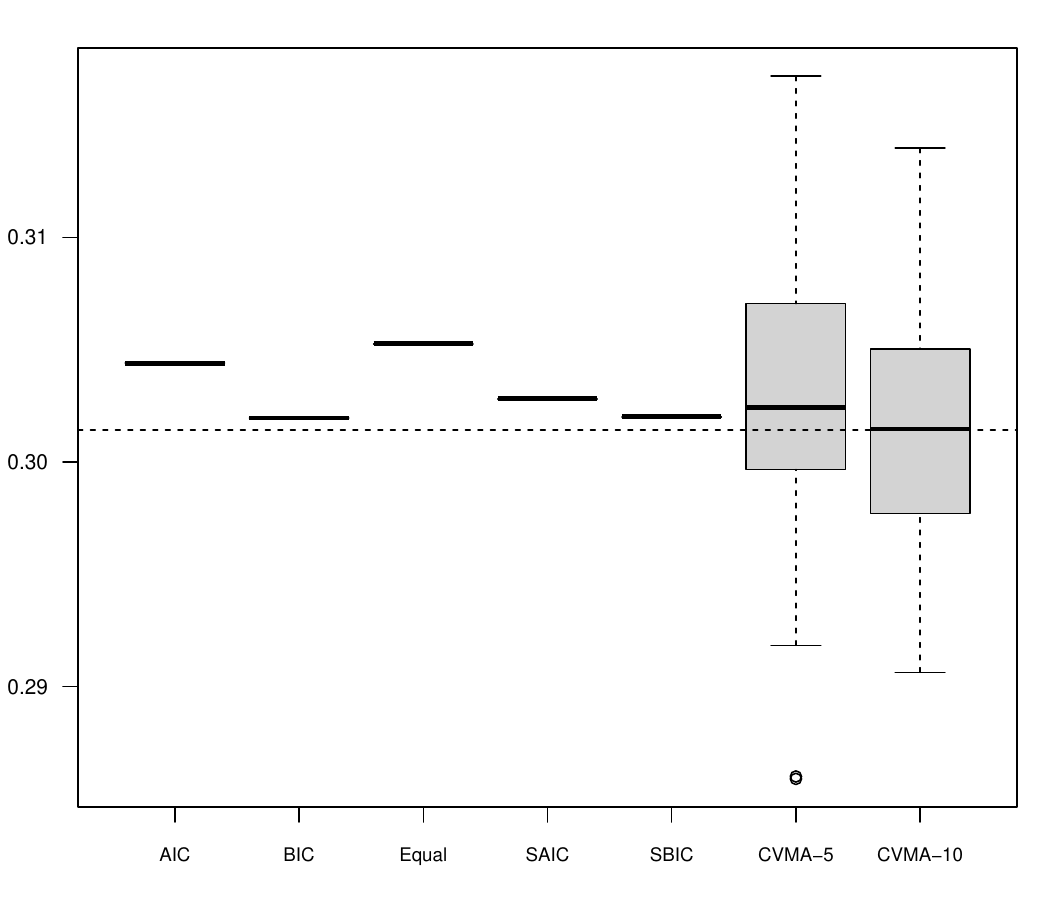}
	\caption{Boxplot of MSPE results for {\it NIR shootout 2002} dataset. The horizontal dashed line indicates the average MSPE of 10-fold CVMA across 100 runs, used as a reference for comparison with other methods.}
	\label{fig:shootout}
\end{figure}

Figure~\ref{fig:shootout} presents the MSPE results for all methods applied to the {\it NIR shootout 2002} data. It is evident from Figure~\ref{fig:shootout} that the 10-fold CVMA estimator yields the minimum MSPE in the average sense, indicating that the proposed procedure is highly effective in prediction. Furthermore, 5-fold CVMA exhibits greater variation than 10-fold CVMA, which is consistent with the typical experience in CV application. Among the other methods, BIC and SBIC produce relatively smaller MSPE values than the other methods. AIC performs poorly in this dataset, and SAIC improves its MSPE, which again demonstrates the effectiveness of the model averaging procedure. Moreover, the large MSPE of equally weighting method indicates its inapplicability in most data analysis practices.

\begin{table}[h]
	\caption{The primary candidate models selected by 10-fold CVMA with their corresponding weights from one run.}
	\label{tab:cvma}
	\centering
	\begin{tabular}{@{}ccccccccccc@{}}
		\toprule
		&$X_1$&$X_2$&$\xi_1$&$\xi_2$&$\xi_3$&$\xi_4$&$\xi_5$&$\xi_6$&$\xi_7$&$\omega_m$\\
		\midrule
        Model 1&\checkmark&0&\checkmark&\checkmark&\checkmark&\checkmark&\checkmark&\checkmark&0&0.7151\\
        Model 2&0&\checkmark&\checkmark&\checkmark&0&0&\checkmark&\checkmark&\checkmark&0.0813\\
        Model 3&\checkmark&0&\checkmark&\checkmark&\checkmark&\checkmark&\checkmark&0&0&0.0808\\
        Model 4&0&\checkmark&\checkmark&\checkmark&\checkmark&\checkmark&0&0&\checkmark&0.0655\\
        Model 5&0&\checkmark&\checkmark&\checkmark&\checkmark&0&\checkmark&\checkmark&0&0.0342\\
        Model 6&0&\checkmark&0&\checkmark&\checkmark&\checkmark&\checkmark&\checkmark&0&0.0180\\
        Model 7&\checkmark&0&0&\checkmark&\checkmark&\checkmark&\checkmark&\checkmark&0&0.0051\\
		\botrule
	\end{tabular}
\end{table}

\begin{figure}[!htb]
	\centering
	\includegraphics[width=\textwidth]{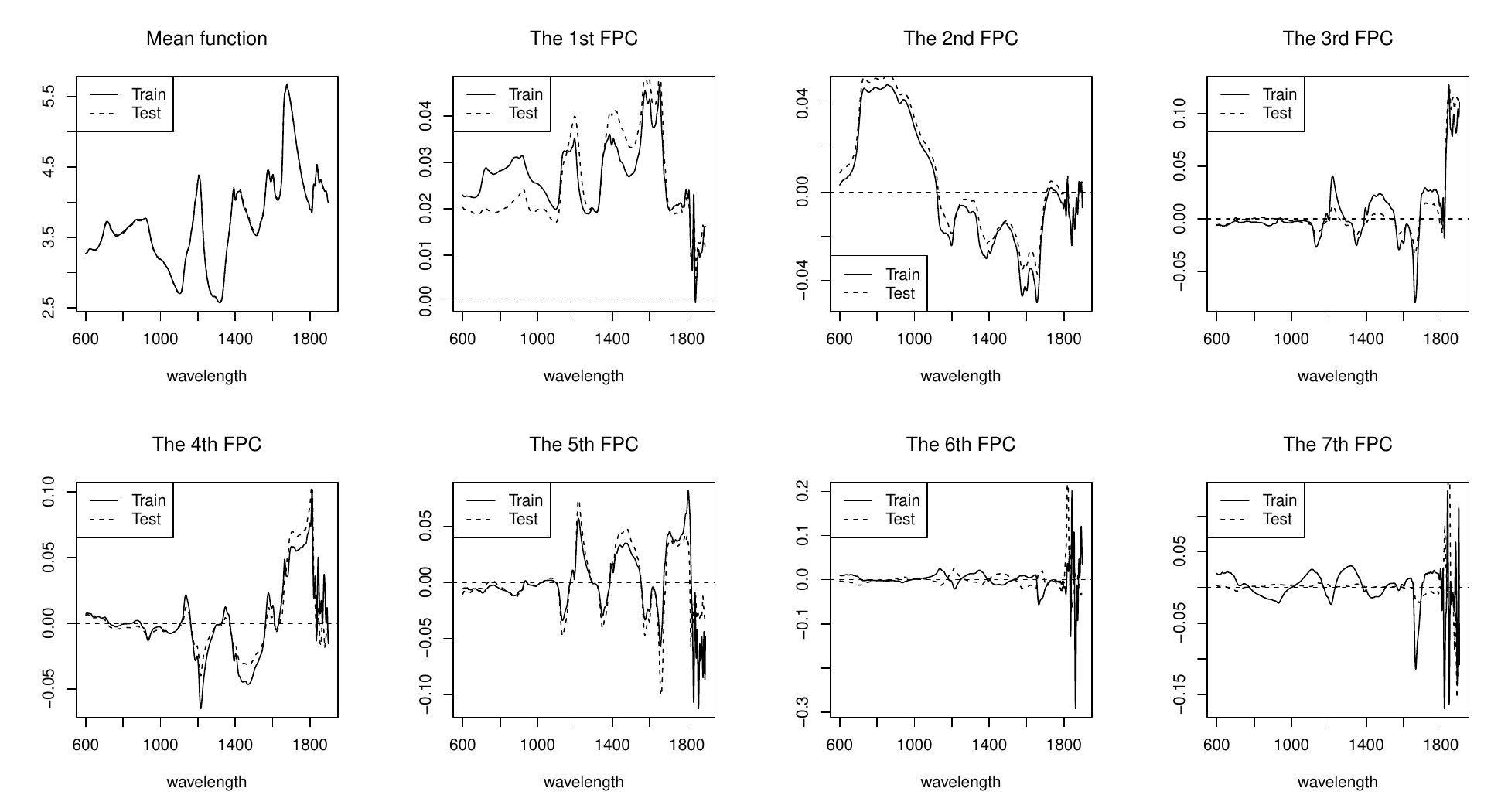}
	\caption{Mean function and leading 7 FPCs of NIR spectra data in training and test subsets.}
	\label{pcplot}
\end{figure}

Furthermore, we randomly picked one result from 100 runs and investigate the detailed performance of the 10-fold CVMA procedure. Table~\ref{tab:cvma} displays the most predictive candidate models selected by 10-fold CVMA, with weights larger than 0.00001 assigned to these models. It is observed that these seven candidate models account for almost all of the proportion. We found that tablet hardness ($X_1$) played an important role in averaging prediction, as it occurred in three candidate models with large weights. Conversely, the information on tablet weight ($X_2$) contributed less to prediction. In terms of the functional variable, $\xi_1$ and $\xi_2$ were frequently used in model averaging with large weights, whereas $\xi_7$ was the least predictive transformed FPC score. Additionally, we presented the mean function and the leading seven FPCs of NIR spectra data in Figure~\ref{pcplot}. Notably, there were differences between the training and the test sets in all these seven FPCs. The trends of the first five FPC curves were basically consistent between the two sets. However, the curves of the sixth and the seventh FPCs showed inconsistent trends, which may partly explain the reason for the less total weights assigned to $\xi_6$ and $\xi_7$ in 10-fold CVMA.

\section{Conclusion and discussion}
This paper investigated a cross-validation model averaging procedure under the framework of partial linear functional additive models. We established the asymptotic optimality of the weight vector selected by a Q-fold cross-validation criterion in the sense of achieving the smallest possible square prediction error loss. Empirical studies demonstrated its superiority or comparability to other methods. However, there are still open questions to be addressed in further research. First, a suitable model averaging estimation for high-dimensional cases may be desired if lots of variables, scalar or functional, are available. \cite{andoli2017} and \cite{zouetal2022} have discussed model averaging procedures for high-dimensional generalized linear models and divergent-dimensional Poisson regression models, respectively, both utilizing Kullback-Leibler loss. Extending these approaches to functional regressions would be of great value. Second, it would be worthwhile to explore consistent model averaging approaches for cases where the correct model exists in the candidate set. \cite{zhangliu2023} has demonstrated the consistency of K-fold model averaging in a quasi-likelihood framework, while several other consistent model averaging estimators in a few non-functional regression models based on Kullback-Leibler loss have been proposed, see \cite{fangetal2022}. This would be an interesting topic for functional regression models. Finally, combining model preparation and model averaging in functional regression modeling is an open question that warrants further investigation. Model preparation can help screen out useful candidate models and hence alleviate difficulties in subsequent analysis. To the best of our knowledge, these three questions have not been discussed in model averaging procedure for functional regression models yet, and deserve further investigation. These avenues of research hold promise for enhancing the prediction performance of functional regression models.

\backmatter

\bmhead{Supplementary information}

The supplementary material provides the proofs of Theorem~\ref{thm1} and additional simulation results.

\bibliography{refmanu}

\end{document}